\documentclass[aip,twocolumn,letterpaper]{revtex4}

\usepackage{fullpage}

\usepackage{xcolor}
\definecolor{ultramarine}{rgb}{0.07, 0.04, 0.56}

\usepackage{graphics}
\usepackage{epsfig}

\usepackage[margin=1.0in]{geometry}

\begin{document}
\title{Use of current-potential patches to obtain fundamental improvements to the coils of magnetic fusion devices}
\author{Allen H Boozer}
\affiliation{Columbia University, New York, NY  10027 \linebreak ahb17@columbia.edu}

\begin{abstract} 
A central issue in the design of tokamaks or stellarators is the coils that produce the external magnetic fields.  The  freedom that remains unstudied in the design of coils is enormous.  This freedom could be quickly studied computationally at low cost with high reliability.  In particular, the space between toroidal field or modular coils that block access to the plasma chamber could be increased by a large factor.  The concept of current-potential patches, which was developed in Todd Elder's thesis, provides a method for separating the study of the feasibility of coils with attractive physics properties from the engineering design of specific coils.   In addition to enhanced accessibility, coils can be designed for increased  plasma-coil separation, insensitivity to coil position errors, and plasma control. \color{black}

 \end{abstract}

\date{\today} 
\maketitle


\section{Introduction}

Unlike many issues that must be addressed before the feasibility of fusion power can be demonstrated, the design of the coils that produce the external magnetic field in toroidal fusion devices has only mathematical subtleties.  Although these issues could be inexpensively addressed by computations, remarkably little has been done \cite{Comp-needs}.  Four issues that are affected by the design of coils are plasma chamber accessibility, plasma-coil separation, insensitivity to coil position errors, and plasma control. 

The development of coil systems that have attractive properties has both physics and engineering aspects, which can to a large extent be separated by the concept of current potential patches.  Peter Merkel showed \cite{Merkel} that an arbitrary current within a toroidal surface can be represented by a current potential.  The current potential is mathematically of two distinct types.  A single-valued part $\kappa(\theta,\varphi)$, which is a periodic function of the poloidal $\theta$ and the toroidal $\varphi$ angles.  This part can be interpreted as the density of magnetic dipoles oriented normal to the toroidal surface  \cite{ITER first wall}\color{black}.  Two non-single-valued parts:  One is proportional to the toroidal angle, has the form $G_{tot}\varphi/2\pi$, and produces the toroidal flux enclosed by the toroidal surface.  The other is proportional to the poloidal angle and produces the poloidal magnetic flux enclosed by the hole in the toroidal surface.  In a toroidal plasma, this part is only relevant for producing a loop  voltage\color{black}---a non-essential part of stellarator design.  The current $G_{tot}$ is the total number of Amperes of current in coils encircling the plasma poloidally, which mathematically may be in one or in many coils.  

Traditionally, the single-valued current potential $\kappa(\theta,\varphi)$ has been Fourier decomposed, but due to the Gibbs phenomenon, a Fourier series representation essentially precludes having localized coils.  A discrete representation is required---patches of current potential on the toroidal surface, which can be represented as dipoles in cells on the surface.  Patches of dipoles oriented normal to an enclosing surface can represent any external non-plasma-encircling coils.  The use of this representation does not imply the final coil set consists of or even contains dipoles.  \color{black}

The feasibility of making changes in the components within the toroidal region enclosed by the coils affects the versatility of fusion experiments and is essential for the maintenance of power plants.  Frequent replacement of internal components in power plants will be required due to limits of materials to neutron fluence, $\sim 10~$MW$\cdot$yr/m$^2$, and the control of tritium during maintenance operations is challenging  \cite{ITER hot Cell: 2017}.   Whatever can be done to remove impediments to plasma-chamber access should be. 

Limits on magnetic field ripple caused by coils that encircle the device poloidally, tokamak toroidal-field coils and stellarator modular coils, prevent easy access to the plasma chamber.  As will be discussed in Section \ref{sec:ripple} and illustrated in Figure \ref{fig:removable-coils}, the number of poloidally-encircling coils could be reduced by the use of localized coils that could be removed along with sections of the chamber walls when access is required.  In principle, only one poloidally-encircling coil is required.  The optimal number requires calculations that have not been made.  

Localized coils that could be removed along with sections of the chamber walls can also be used to minimize limitations on the plasma-coil separation, to mitigate the effects of coil position errors, and provide maximal external magnetic field control.  The benefits of a general coil set offers for plasma control and for the mitigation of the effects of construction errors was extensively discussed in \cite{Comp-needs}, and that discussion will not be repeated here.

\begin{figure}
\centerline{ \includegraphics[width=1.5 in]{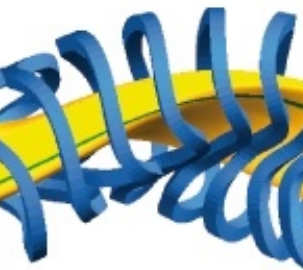}}
\caption{Modular coils are shaped toroidal field coils that allow the difficulties of helical coils to circumvented.  Just as toroidal field coils in tokamaks, modulars must be closely packed around a stellarator to avoid an unacceptable ripple in the magnetic field strength.  Non-planar shaping of the modulars that increases linearly with their radius is required to produce elliptical shaping of the magnetic surfaces and quadratically to produce triangularity.  Axis torsion, which comes from horizontal and vertical fields, can be produced by the placement of planar modular coils.  This is part of a figure produced by the Max-Planck Institut f\"ur Plasmaphysik, which can be used without requesting permission.  } 
\label{fig:modualar}
\end{figure}

The paradigm of stellarator coil design has become  shaped toroidal field coils, called modular coils, Figure \ref{fig:modualar}.  They were introduced in 1972 by Wobig and Rehker \cite{Modulars:1972} and can be efficiently designed using the FOCUS code \cite{Focus code: 2017}.  Modular coils have three fundamental limitations \cite{Comp-needs}:  (i) There is a severely limiting tradeoff between access to the of plasma chamber and the ripple in the magnetic field strength.  (ii)  Power plants require a large plasma-coil separation and stellarator design is eased when a large separation is possible.  The required shaping of modular coils increases the larger the coil, $b$, to plasma, $a$, radius ratio, approximately as $(b/a)^{m-1}$ with $m$ the poloidal mode number of the field or the plasma shape.  Limits on the shaping may set the maximum plasma-coil separation at a smaller value than the separation that would be possible with a more general coil type.  (iii) The modular coils themselves offer little flexibility for plasma control.  


\section{Representation of an Arbitrary External Magnetic Field \label{Sec: dipole rep}}

Peter Merkel's method of representing a general external magnetic field  \cite{Merkel} was shown in \cite{ITER first wall} to imply any magnetic field within a region enclosed by a toroidal surface that is produced by coils external to that region could be produced by one more coils encircling the plasma poloidally and carrying a total current in Amperes $G_{tot}$ plus a set of dipoles oriented normal to the enclosing surface that have a density per area given by a periodic function $\kappa(\theta,\varphi)$.  The dipole moment of a small area on the surface is $\hat{n}\int \kappa da$, where $\hat{n}$ is the outward normal to the surface.  Dipoles oriented normal to the surface that densely cover an enclosing surface give a general representation of any externally produced external magnetic field; no additional generality is achieved by considering dipoles that have other than an orientation along $\hat{n}$. 

The proof starts by noting that the general magnetic field within a region enclosed by a toroidal surface that is produced by coils external to that region is curl and divergence free.  Therefore, it can be represented as $\vec{B}_x=\vec{\nabla}\phi_x$ with $\nabla^2\phi_x=0$.  

Positions on a toroidal surface are described by two angles, a toroidal angle $\varphi$, which advances by $2\pi$ each toroidal circuit, and a poloidal angle $\theta$. Each of these angles is arbitrary in the sense that $\varphi+\omega_\varphi(\theta,\varphi)$ and $\theta+\omega_\theta(\theta,\varphi)$, where $\omega_\varphi$ and $\omega_\theta$ are periodic functions, are valid choices. 

The general form of the magnetic field in the enclosed region is 
\begin{equation}
\phi_x = \frac{\mu_0 G_{tot}}{2\pi}\varphi - \frac{\mu_0 I_{tot}}{2\pi}\theta + \phi_p, \label{B-pot}
\end{equation}
where $\phi_p$ is a periodic function of $\theta$ and $\varphi$.  The two angles need not satisfy Laplace's equation, but the function defined by $\nabla^2\varphi$ will be periodic and can be absorbed into $\phi_p$.   

The current $G_{tot}$ is proportional to the toroidal magnetic flux in the enclosed region because the toroidal loop integral of the magnetic field equals $\mu_0 \vec{j}$ integrated over the area enclosed by the loop $\oint \vec{B}\cdot d \vec{x}_\varphi= \mu_0 \int \vec{j}\cdot d\vec{a}$.  Similarly the magnetic flux passing through the central hole of the torus is proportional to $I_{tot}$.  Within the enclosed toroidal region, the flux coming through the central hole is only relevant when it varies, which produces a loop voltage.  Otherwise, the field in the enclosed region can be produced by the periodic part of the potential plus the toroidal flux produced by $G_{tot}$.

The magnetic field outside of the shell has the same form at Equation (\ref{B-pot}), and currents outside the shell can contribute to the normal magnetic field on the shell $\vec{B}\cdot\hat{n}$.  This normal field gives a Neumann boundary condition on the solution for $\phi_x$ in the region enclosed by the shell;  $\vec{\nabla}\cdot \vec{B}=0$ ensures $\vec{B}\cdot\hat{n}$ cannot change as a thin shell is crossed.

Although a Neumann boundary condition defines $\phi_x$ within the enclosed region to within an irrelevant constant, the current flowing in the shell produces a jump in $\phi_x$ and hence affects $\vec{B}\cdot\hat{n}$ on the shell.  Equation (\ref{kappa eq}), which is derived below, shows that as the shell is crossed $\big(\phi_x\big)_{out} - \big(\phi_x\big)_{in} = \mu_0\kappa_{tot}$, where $\kappa_{tot}$ is the total current potential.  This current potential is the sum of a periodic part $\kappa(\theta,\varphi)$ and a non-periodic part given by the net poloidal and toroidal currents in the shell.  

When explicit coils define the net poloidal and toroidal currents, then only the periodic part of the current potential needs to be retained, and that part will be shown in Section \ref{Sec:moment} to be representable by a sufficiently fine mesh on the enclosing surface with each mesh cell having a dipole moment normal to the surface with an amplitude given by the integral of $\kappa$ over the cell, $\vec{m} =\hat{n}\int \kappa da$.


\subsection{Coordinates near a surface}

This section proves the unit normal $\hat{n}$ to a smooth toroidal surface is curl free and can be used to define positions in the vicinity of that surface.  This is needed to obtain the general relation between a current in a thin shell and the jump in $\phi_x$ across the shell.  
 
A smooth toroidal surface can be given using arbitrary poloidal and toroidal angles using a function $\vec{x}_s(\theta,\varphi)$.    Positions near that surface can be defined in $(r,\theta,\varphi)$ coordinates in which $r=b$ is the surface as
\begin{eqnarray}
\vec{x}(r,\theta,\varphi) &=& \vec{x}_s(\theta,\varphi) + (r-b)\hat{n}.  \label{coord-system}\\
\hat{n} &\equiv& \frac{ \frac{\partial \vec{x}_s}{\partial\theta}\times \frac{\partial \vec{x}_s}{\partial\varphi}}{\left|  \frac{\partial \vec{x}_s}{\partial\theta}\times \frac{\partial \vec{x}_s}{\partial\varphi}  \right|} \nonumber\\
&=& \vec{\nabla}r. \label{grad r}
\end{eqnarray}
The definition of the unit normal and its relation to the gradient of the radial coordinate, $\hat{n}=\vec{\nabla}r$, follow from the theory of general three-dimensional coordinate systems, which is derived in less than two pages in the Appendix to \cite{Boozer:RMP}.  The Jacobian of $(r,\theta,\varphi)$ coordinates  is 
\begin{eqnarray}
\mathcal{J}&\equiv& \frac{\partial \vec{x}}{\partial r}\cdot\left( \frac{\partial \vec{x}}{\partial\theta}\times \frac{\partial \vec{x}}{\partial\varphi}\right)  \\
&=& \left|  \frac{\partial \vec{x}_s}{\partial\theta}\times \frac{\partial \vec{x}_s}{\partial\varphi}  \right| \hspace{0.2in} \mbox{at $r=b$.} \hspace{0.2in}
\end{eqnarray}

\subsection{The general surface current}

Any divergence-free vector can be written as a cross product of gradients.  Any divergence-free current density in a toroidal shell can be written as $\vec{j} = \vec{\nabla}\Big(\kappa_{tot} \delta(r-b)\Big)\times \vec{\nabla}r$, where $\kappa_{tot}(\theta,\varphi)$ is the total current potential.  Note the derivative of the delta function, $ \delta(r-b)$ is made irrelevant by crossing it with $\vec{\nabla}r$.  Since $\vec{\nabla}r=\hat{n}$, the current density is $\vec{j}=\vec{\nabla} \times \big(\kappa_{tot}\delta(r-b)\hat{n}\big)$. 

The surface current in the shell is defined by an integral across the shell,
\begin{eqnarray}
\vec{J} &\equiv& \int \vec{j} dr \\
&=& \vec{\nabla}\kappa \times \hat{n}.
\end{eqnarray}

Ampere's law for a surface current can be written as $\vec{\nabla}\times\big(\vec{B} - \mu_0\kappa_{tot}\delta(r-b)\hat{n}\big)=0$, which implies there is a potential $\phi_A$ with $\vec{\nabla}\phi_A=\vec{\nabla}\phi_B -\kappa_{tot}\delta(r-b)\hat{n}$.  When the magnetic field is curl free just outside of the shell, the relation between the scalar potential for the magnetic field on the two sides of the enclosing surface is obtained by integrating with respect to $r$ across the delta function; 
\begin{equation}
\big(\phi_x\big)_{out} - \big(\phi_x\big)_{in} = \mu_0\kappa_{tot}. \label{kappa eq}
\end{equation}
There is a sign error in Equation (11) in \cite{ITER first wall}.

The divergence-free nature of the magnetic field implies $\vec{B}\cdot\hat{n}$ cannot change across a shell described by a delta function, so
\begin{equation}
( \hat{n}\cdot\vec{\nabla}\phi_p)_{out} = ( \hat{n}\cdot\vec{\nabla}\phi_p)_{in}. \label{Norm eq}
\end{equation}


\subsection{Magnetic moment associated with a current potential \label{Sec:moment}}

The magnetic field produced by a single-valued current potential can be calculated by dividing the toroidal surface into small cells, so small that $\hat{n}$ does not change significantly across the cell.  The current potential in each cell will be shown to define a magnetic dipole with a moment $\vec{m}=\hat{n}\int_{cell} \kappa da$. 

The magnetic moment of a region of space is $\vec{m} = \frac{1}{2} \int \vec{x}\times\vec{j} d^3x$.  The current density of the single-valued part of the current potential is $\vec{j}=  \delta(r-b) \vec{\nabla}\kappa \times \hat{n}$.  Using $(r,y,z)$ as Cartesian coordinates to describe a single cell, the magnetic moment is
\begin{eqnarray}
\vec{m} &=& \frac{1}{2} \int \vec{x}\times \big(\delta(r-b) \vec{\nabla}\kappa \times \hat{n}\big) dr dy dz \\
&=& \int \frac{\delta(r-b)}{2}   \big\{(\vec{x}\cdot\hat{n})\vec{\nabla}\kappa  - (\vec{x}\cdot\vec{\nabla}\kappa)\hat{n} \big\} dr dy dz \hspace{0.2in}\\
&=&-\frac{\hat{n}}{2} \int (\vec{x}_s \cdot \vec{\nabla}\kappa)dydz\\
&=&-\frac{\hat{n}}{2} \int \{y \frac{\partial \kappa}{\partial y} + z \frac{\partial \kappa}{\partial z}\} dydz\\
&=&\hat{n} \int \left\{\kappa - \frac{1}{2}\left(\frac{\partial y \kappa}{\partial y}+\frac{\partial z \kappa}{\partial z}\right)\right\} dydz\\
&=& \hat{n}\int \kappa dydz
\end{eqnarray}
since $\kappa$ is non-zero only within the cell


\section{Use of the dipole representation}
\color{black}

Peter Merkel \cite{Merkel} used the current potential in Fourier decomposed form to develop a design for the coils of the W7-X stellarator; an improved version of the method has been given by Landreman \cite{Landreman:2017}.

A Fourier series cannot accurately represent a surface current that is zero in large regions of the surface.  Surface currents that are non-zero only in certain locations require a discrete representation of $\kappa$.   As shown in Section \ref{Sec: dipole rep}, \color{black} this can be is achieved \cite{Comp-needs} by dividing the current-potential surface into cells.  Each cell has a magnetic dipole in its center, oriented normal to the surface.  The magnitude of the dipole moment of the $\ell^{th}$ cell is the current potential integrated over its area, $d_\ell=\int \kappa da_\ell$.

The index $\ell$ is only subtilely related to the distance between dipoles.  To determine which dipoles are nearest neighbors two indices are needed, $d_{\mu\nu}$.  The Greek $\mu$ is a poloidal and $\nu$ a toroidal index with $\ell$ the prescribed method for numbering dipoles given by $\mu$ and $\nu$.  As will be discussed in Section \ref{sec:forces}, the gradient of current potential cannot be too large and have acceptable forces.  This gradient can be reduced by defining a smoothed dipole distribution $<d>_{\mu\nu}$ by 
\begin{equation}
<d>_{\mu\nu} \equiv \frac{d_{\mu+1,\nu} + d_{\mu-1,\nu} + d_{\mu,\nu+1} + d_{\mu,\nu-1}  }{4} \label{smoothed d}
\end{equation}
and requiring the normal magnetic field be adequately produced by $<d>_\ell$.

\begin{figure}
\centerline{ \includegraphics[width=2.5 in]{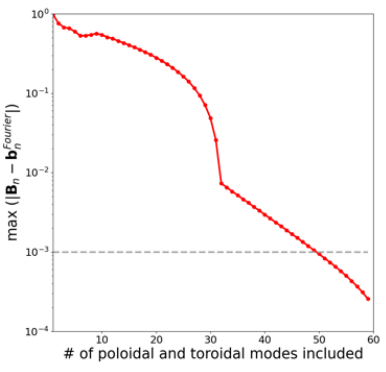}}
\caption{The maximum error in a Fourier decomposition of the normal component of axisymmetric magnetic field on the outer plasma surface of the HSX stellarator is given versus the number of Fourier components that are retained.  The estimated requirement is a maximum error  of $10^{-3}$ of the magnetic field .  This is Figure 1 in Elder and Boozer \cite{Elder:2024}. } 
\label{fig:B-n Fourier}
\end{figure}

The required normal magnetic field $(\vec{B}\cdot\hat{n})_{req}$ \color{black} to an arbitrary surface that is enclosed by the current-potential surface can be Fourier decomposed with $b_k$ the $k^{th}$ Fourier component.  Fourier series of analytic functions converge exponentially, so only a moderate number, $K\approx50$, terms $b_k$ is required to produce the normal field with the required accuracy, Figure \ref{fig:B-n Fourier}.  This means that the maximum error in the normal field anywhere on the surface is less than some fraction of the total magnetic field, such as a part in a thousand.  

Using the expression for the magnetic field due to a dipole, the matrix $\tensor{M}$, which relates the dipole matrix vector $\vec{d}$ with components $d_\ell$ and the vector $\vec{b}$ with components $b_k$, gives 
\begin{equation} \vec{b} = \tensor{M}\cdot \vec{d}. \label{mutual} \end{equation}

The concept of a gridded current potential is related to that of permanent magnets \cite{Zhu et at,Hammond:2024} or a large number of simple electromagnets \cite{Thea Energy:2023} to produce part of the magnetic field of a stellarator.  Kaptanoglu et al \cite{Kaptanoglu:2024a} have used volume elements (voxels) of electric current to define stellarator coils.  Unlike those concepts, a gridded current potential is purely a computational method by which physicists can study strategies for coils.   Once an attractive $\kappa$ is determined, which includes acceptable forces, Section \ref{sec:forces}, many coil choices are possible.  Which choice is optimal is an issue of engineering.

The singular values of a singular value decomposition (SVD) of $\tensor{M}$ and their associated left and right singular vectors determine what normal magnetic field distributions can be efficiently driven by which distributions of surface current.  The condition number of a singular left or right eigenvector is the ratio of the largest singular value of $\tensor{M}$ to the singular value associated with that eigenvector.   Spatially-constant horizontal and vertical magnetic fields do not decay with distance from the coils that produce them and can be used to normalize the efficiency. The only magnetic fields that can be feasibly produced are those with a sufficiently small condition number---approximately five.  The number of distributions of dipoles that are consistent with a condition-number limit is far larger than the condition number itself for physically relevant condition numbers.  See Figure 2 in Elder and Boozer \cite{Elder:2024}.

The concept of efficient magnetic fields was introduced by Boozer in Section V.D.1 of a 2004 review of magnetic confinement \cite{Boozer:RMP}.  Landreman and coauthors have published two other articles \cite{Landreman:eff-B,Landreman:field-gradient} on the distance of feasible separation between the coils and the plasma.

The condition number of a singular-eigenvector of a dipole distribution, which has a magnitude $d_\lambda$, is denoted as $C_\lambda$.  More important than the maximum condition number $C_\lambda$ that is included in a solution is the effective condition number, 
\begin{equation}
C_{eff}\equiv \sqrt{ \frac{\sum_\lambda (C_\lambda d_\lambda)^2}{\sum_\lambda d_\lambda^2}}. \label{eff-C}
\end{equation}
The inclusion of a dipole distribution that inefficiently produces a required normal-field distribution is not important when the required amplitude $d_\lambda$ is sufficiently small.

Not all errors in the normal field degrade the physical properties of the plasma equilibrium at a moderate amplitude \cite{Ku-Boozer:2010,Boozer-Ku:control2011}, and a penalty should not be placed on the coils when such fields are driven at a low amplitude.  A sensitivity study of the plasma equilibrium is required to take advantage of what can be a large reduction in the number of Fourier components that need to be produced.  Once a set of $L$ dipoles is chosen to produce the required field, a check should be made of the quality of the resulting plasma equilibrium.  

As in other applications in which an SVD analysis is used to achieve a smooth fitting, the number of free parameters, the number of dipoles $L$, should be large compared to the number points to be fit, the number of Fourier coefficients $K$.


\section{Required coil coverage}

\begin{figure}
\centerline{ \includegraphics[width=2.5 in]{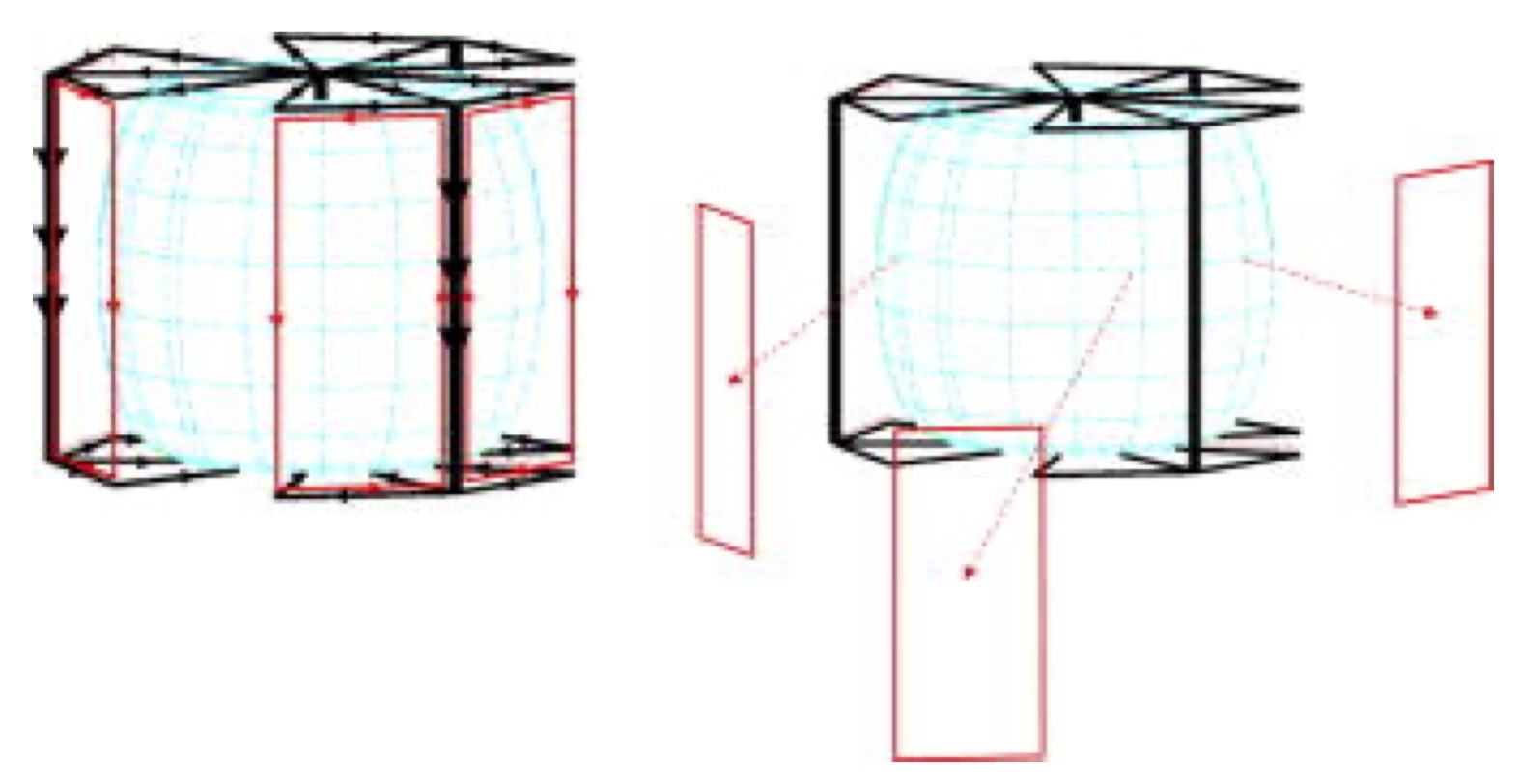}}
\caption{Non-plasma encircling coils mounted on removable wall sections can be used to increase the accessibility of fusion plasmas.  This was Figure (4) from the 2009 publication \cite{Boozer:2009} and illustrated a way to enhance plasma access in tokamaks.   In this case, the distance between encircling toroidal field coils was increased by a factor of three. } 
\label{fig:removable-coils}
\end{figure}

The required access to the plasma chamber is of two types: port and maintenance.  Port access means an open channel exists for diagnostics, particle injection, etc.  Maintenance access means a section of the chamber wall, together with any coils attached to it, can be removed, which allows large internal components to be replaced, Reference \cite{Boozer:2009} and Figure \ref{fig:removable-coils},  It is important to study how the coils that produce the toroidal magnetic field, in tokamaks as well as stellarators, can be made consistent with easier access to the plasma chamber.  Both types of chamber access become better when the fraction of the wall covered by coils is reduced and when one can locate the coils away from regions where an access conflict would exist.  

Although there is no requirement that the cells on the current-potential surface  have equal-area, this will be assumed to be to simplify the discussion. 

\subsection{Todd Elder's analysis}

As part of his doctoral research with Allen Boozer, Todd Elder studied \cite{Elder:2024} how much of the current potential surface needs to be covered with dipoles to produce the external field required for the HSX stellarator \cite{HSX} in the presence of an axisymmetric toroidal field.  He found that the HSX magnetic surfaces could be accurately recovered by only a 22 \% coverage of the current potential surface.  

\begin{figure}
\centerline{ \includegraphics[width=2.5 in]{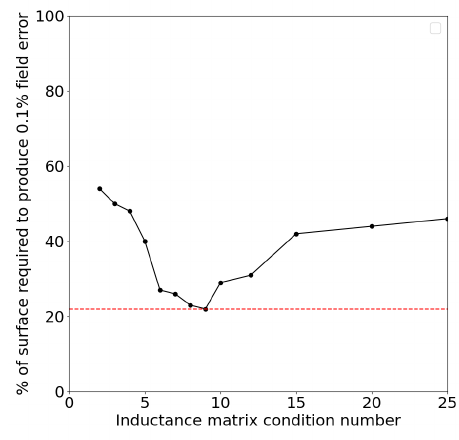}}
\caption{By eliminating dipoles of low amplitude from the matrix $\tensor{M}$, the fraction of the current-potential surface covered by dipoles can be reduced until the required HSX field can no longer be fit to the chosen maximum error of $0.1\%$.  For small condition numbers of the highest singular value allowed, the required coverage drops as one would expect.  For larger condition numbers the fit becomes worse, because singular vectors with high condition numbers contain dipoles that buck each other out.  This was Figure 4 in Elder and Boozer \cite{Elder:2024}.} 
\label{Area-cond}
\end{figure}

\begin{figure}
\centerline{ \includegraphics[width=3.2 in]{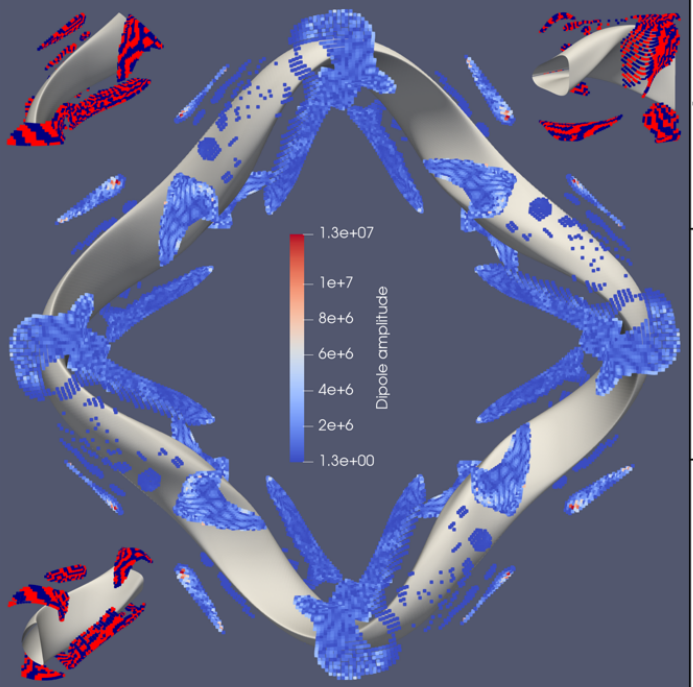}}
\caption{The location of dipoles when only 22\% of the surface is covered. The color bar shows the unsigned strength of the dipoles, which are predominately blue.  The three corner diagrams give the signs of the dipoles in red and black. This was part of Figure 3 in Elder and Boozer \cite{Elder:2024}} 
\label{Patch-cov}
\end{figure}

The strategy that Elder adopted for reducing the coverage of the wall was to start with complete coverage but remove those dipoles that had a low magnitude.  While doing this, he determined, Figure \ref{Area-cond},  how the largest condition number $C_\lambda$ among the included fields increased in order to achieve a maximum error of 0.1\% of the total magnetic field.  He found that there was a minimum area-fraction of approximately 20\%, and for higher condition numbers the required area actually increased---presumably due to distributions with a high $C_\lambda$ including bucking dipole distributions.  The distribution of dipoles when the coverage was $\approx22\%$ is shown in Figure  \ref{Patch-cov}.

Elder's work is very important.  As is typical, important results tend to raise more questions than were answered, and a different strategy might give an even more optimistic result.


\subsection{Importance of each dipole}

\color{black}

The strategic question is how should the retained dipoles for a minimum-area solution be determined.  There are two reasons dipoles can have a large amplitude: (i) They are important for producing the required magnetic field.  (ii) They drive a relevant magnetic field but only inefficiently.  These two causes can explain why Figure \ref{Area-cond} has a minimum.



An optimal minimum-area solution requires a method for picking the dipoles based on their importance.  Each magnetic dipole $d_{\mu\nu}$ produces a specific normal field $(\vec{B}\cdot\hat{n})_{\mu\nu}$ on the surface on which the required normal field is $(\vec{B}\cdot\hat{n})_{req}$.  The importance of the dipole $d_{\mu\nu}$ is  
\begin{equation}
\mathcal{I}_{\mu\nu}\equiv\frac{|\oint (\vec{B}\cdot\hat{n})_{\mu\nu} (\vec{B}\cdot\hat{n})_{req} d\theta d\varphi |}{|d_{\mu\nu}|}.
\end{equation}
When the dipole grid is sufficiently fine to accurately represent the required magnetic field, $(\vec{B}\cdot\hat{n})_{req}$, the importance measure $\mathcal{I}_{\mu\nu}$ varies slowly with $\mu$ and $\nu$.

\subsection{Obtaining compact regions of non-zero dipoles}

\color{black}

Most distributions of dipoles that are determined by an SVD decomposition of $\tensor{M}$ have no effect on the $K$ retained Fourier components when the number $L$ of retained dipoles greatly exceeds $K$.  Mathematics implies that the number of non-zero singular values cannot be larger than the smaller of $K$ or $L$.  Removing the dipole distributions that are associated with a singular value of zero, which is most of the distributions, sounds like an obvious way to proceed.  Unfortunately, this need not give compact regions of non-zero dipoles. 

It seems preferable to have the retained dipoles lie in a small number of contiguous regions.  The matrix $\tensor{M}$ places no preference on the minimization of the number of contiguous regions.  

 A contiguous region consists of dipoles $d_{\mu\nu}$ in a continuous range of $\mu$ and $\nu$.  The slow variation of $\mathcal{I}_{\mu\nu}$ with respect to $\mu$ and $\nu$ gives an obvious way to choose important contiguous regions---regions in which $\mathcal{I}_{\mu\nu}>c_{\mathcal{I}}$ for a chosen constant $c_{\mathcal{I}}$. 

Any choice of the constant $c_{\mathcal{I}}$ gives a definite number of dipoles that are included in contiguous regions, and therefore a definite fractional coverage of the dipole surface.  As $c_{\mathcal{I}}$ is made smaller, more of the dipole surface is covered and the better the required normal magnetic field should be fit with an acceptable condition number.  

Some regions defined by a smaller $c_{\mathcal{I}}$ than that used for the primary regions may give a more attractive set of compact coil regions.  This can be studied by calculating the magnetic field that has a relatively large $c_{\mathcal{I}}$ and small condition number that nonetheless produces most of the required normal field.  The residual required-normal-field can then be fit by determining which dipoles have the greatest importance in fitting the residual, which means defining an $\mathcal{I}^{(1)}_{\mu\nu}$.   Contiguous regions can be chosen with  $\mathcal{I}^{(1)}_{\mu\nu}>c^{(1)}_{\mathcal{I}}$ for some $c^{(1)}_{\mathcal{I}}$, which is smaller than $c_{\mathcal{I}}$.  This iterative procedure can be explored using as many steps as beneficial.

A related optimization procedure has been used by Kaptanoglu, Conlin, and  Landreman to design stellarators using permanent magnets \cite{Kaptanoglu:2024b}.  Here the dipoles are purely a technique of discretizing the current potential for a general study of what features of external coils are possible, not the design of the external coils or magnets themselves.

\color{black}


\section{Forces \label{sec:forces} }

The feasibility of a solution is in large part determined by the forces on the coils.  The $\vec{j}\times\vec{B}$ force integrated across a shell that has a current potential $\kappa$ is a force per unit area.  This force is given by Equation (17) of Reference \cite{ITER first wall},
\begin{eqnarray}
\vec{F}_\kappa = \left(\frac{\vec{B}_+ +\vec{B}_-}{2} \cdot\vec{\nabla}\kappa \right)\hat{n} -\left(\vec{B}\cdot\hat{n}\right)\vec{\nabla}\kappa. \label{shell force}
\end{eqnarray}
The first term is the force normal to the surface on which the current potential is located and the second term is the force tangential to the surface.  $(\vec{B}_+ +\vec{B}_-)/2$ is the magnetic field spatially averaged across a thin shell which carries the surface current, and $\vec{B}\cdot\hat{n}$ is the normal magnetic field to the shell.  The magnitude of this force per unit area should be compared to the total force per unit area that can be exerted by the magnetic field, $B^2/2\mu_0$, to assess its implications.  The calculation of the magnetic field at the location of a current potential is subtle but an accurate method has been developed by Malhotra, Cerfon,  O'Neil, and Toler  \cite{Cerfon}.   Far simpler estimates of $(\vec{B}_+ +\vec{B}_-)/2$ and $\vec{B}\cdot\hat{n}$ could be used to determine the consistency of the forces on current-potential patches with their feasibility.   For example, while studying ripple reduction in a tokamak, the field $\vec{B}$ could be approximated as the desired axisymmetric toroidal field.\color{black}

However $\vec{B}$ is calculated, acceptable forces imply that the derivative of $\kappa$ cannot be extremely large.  That is the values of $d_{\mu\nu}$ should be similar for dipoles that are close to each other. Using the smoothed $<d>_{\mu\nu}$ of  Equation (\ref{smoothed d}) to define the magnitude of the retained dipoles mitigates large gradients.


Gaussian smoothing,  as described in Appendix \ref{Gaussian smoothing} gives a $\bar{\kappa}$ that is an analytic function of position.  Additional smoothing can reduce the forces, \color{black} and the smoothing scale $\sigma$ may be chosen to do that as opposed to the choice $\sigma=\sqrt{a_c}$, which optimizes the representation of variations in current potential.


\section{Analytic model of toroidal ripple} \label{sec:ripple}

Figure \ref{fig:removable-coils} illustrates a method of reducing the number of poloidally encircling coils by a factor of three.  An analytic model of a cylindrical solenoid provides intuitive understanding of the general method.  

\begin{figure}
\centerline{ \includegraphics[width=3.2 in]{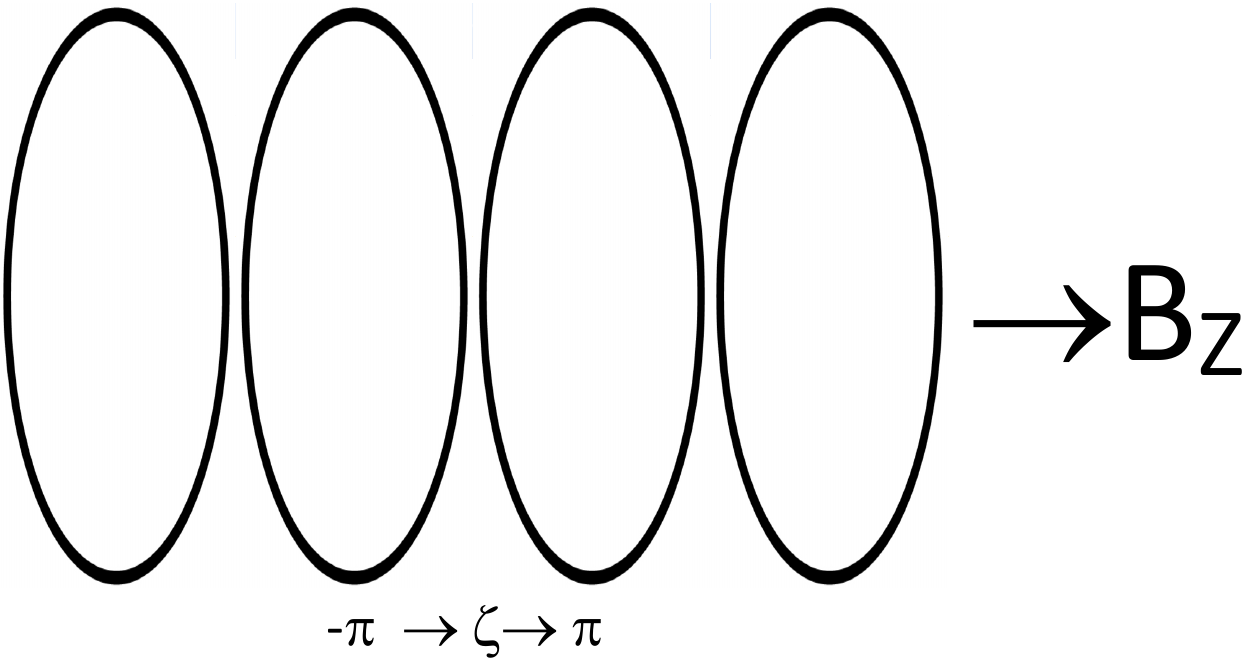}}
\caption{A circular solenoid that produces a $z$ directed magnetic field has circular coils separated by a distance $2\pi/k_c$ each carrying a current $G_c$.  Letting $\zeta\equiv k_cz$, a ripple free $B_z$ is obtained as $k_c\rightarrow\infty$ with $G_ck_c$ held constant.  Between a particular pair of neighboring coils, $\zeta$ goes from $-\pi$ to $\pi$ and the non-single valued current potential in the $k_c\rightarrow\infty$ limit is $G_c\zeta/2\pi$. }
\label{Solenoid}
\end{figure}

The production of the external magnetic field for the a stellarator using modular coils has ripple issues related to those of the toroidal field coils of a tokamak.  But, the optimization of the steady-state external field in a stellarator requires a non-axisymmetric field as well as a strong toroidal magnetic field, which gives far more freedom and subtlety to finding optimal coils.  Tilting planar toroidal field coils can efficiently produce the $m=1$ components of the non-axisymmetric field.  The current-potential patches used to produce the non-axisymmetric fields can also be used to minimize the ripple.  The toroidicity of both tokamaks and stellarators makes the ripple issue more difficult on the out-board than on the in-board side of the machine as is implied by Figure \ref{fig:removable-coils}.  Nevertheless, useful intuition is obtained from an analytic model of a cylindrical solenoid.

The analytic solenoid has circular coils, each carrying a current $G_c$, that are separated by a distance $2\pi/k_c$ in the $z$ direction, Figure \ref{Solenoid}.  As the separation goes to zero, $k_c\rightarrow\infty$, these coils produce a $z$-directed magnetic field, which is the modeled toroidal magnetic field.  The non-single valued current potential in the limit $k_c\rightarrow\infty$ with $G_ck_c$ held constant is $(G_ck_c)z/2\pi$.  The angle $\zeta\equiv k_cz$, which runs from $-\pi$ to $\pi$ between any particular pair of neighboring coils, is used to describe the region between the coils.  The non-single valued part of the current potential is $G_c\zeta/2\pi$ in the limit  $k_c\rightarrow\infty$ with $G_ck_c$ held constant.

The variable $\zeta$ can be Fourier decomposed over the range $\pi > \zeta > -\pi$.  When $n$ is an integer, the integral $\int_0^\pi \zeta \sin(n\zeta) d\zeta= - (-1)^n \pi/n$, so the Fourier  series is
 \begin{equation}
 \zeta = - 2 \sum_{n=1}^\infty \frac{(-1)^n}{n} \sin n\zeta \mbox{   when  } \pi\geq \zeta\geq-\pi.
 \end{equation}
 This series gives zero at $\zeta=\pm\pi$, and as $\zeta$ approaches $\pm \pi$ an accurate representation requires $n\rightarrow\infty$.  This Fourier representation has no validity outside the $|\zeta|<\pi$ range.
 
  The interpretation is that a localized coil carrying a current $G_c$ encircling the plasma poloidally is required where  $|\zeta|\rightarrow\pi$.  The effective current potential when there are no coils between the encircling coils is
 \begin{eqnarray}
 \kappa_{eff} &=& \frac{G_c}{2\pi }\Big(\zeta+ 2\sum_{n=1}^\infty \frac{(-1)^n}{n} \sin n\zeta \Big).
 \end{eqnarray}
 In other words, the resulting magnetic field would have no ripple if a single valued current potential were supplied between the coils with
  \begin{eqnarray}
 \kappa &\equiv& \sum_{n=1}^\infty \kappa_n \sin n\zeta \\
 \kappa_n &\equiv& - \frac{G_c}{\pi }\frac{(-1)^n}{n}.
 \end{eqnarray}
 The ripple that results from having partial or no cancelling current potential can be calculated using the missing single-valued current potential $\kappa$.
 
  Let $\phi$ be the magnetic scalar potential, so $\vec{B}=\vec{\nabla}\phi$.  In the region enclosed by the coils $\nabla^2\phi=0$,  and $\phi$ is determined by Equations (\ref{kappa eq}) and (\ref{Norm eq}) with a single-valued current potential.
  
The solution to the Laplacian $\nabla^2\phi=0$ \color{black} in a cylinder uses modified Bessel functions of the first and second kind:
\begin{eqnarray}
\frac{\phi_{in}}{\mu_0} &=&  \frac{G_c\zeta}{2\pi }  +  \sum_{n=1}^\infty \frac{\phi^{(in)}_n}{\mu_0} I_0(nk_c r)  \sin(n \zeta); \\
\frac{\phi_{out}}{\mu_0} &=&  \sum_{m=1}^\infty \frac{\phi^{(out)}_n}{\mu_0} K_0(nk_c r),  \sin(n\zeta). 
\end{eqnarray} \vspace{0.2in}

Since $dI_0/dx=I_1(x)$ and $dK_0/dx=-K_1(x)$, Equations (\ref{kappa eq}) and \ref{Norm eq} imply
\begin{eqnarray}
\kappa_n &=& - \frac{\phi^{(in)}_n }{\mu_0} \mathcal{I}_0(nk_cb) \mbox{  with   } \\
\mathcal{I}_0(x) &\equiv& I_0(x) +K_0(x) \frac{I_1(x))}{K_1(x)}, \mbox{   and   } \\
\frac{\phi_{in}}{\mu_0} &=& \frac{G_c \zeta}{2\pi }  + \sum_{n=1}^\infty \kappa_n \frac{I_0(nk_c r)}{\mathcal{I}_0(nk_cb)}  \sin(n \zeta). \hspace{0.2in}
\end{eqnarray}
In the absence of a current potential to balance the ripple, the magnetic field in the interior is given by
\begin{eqnarray}
\frac{\phi_{in}}{\mu_0G_c} &=&  \frac{\zeta}{2\pi } + \sum_{n=1}^\infty \frac{(-1)^n}{n\pi} \frac{I_0(nk_c r)}{\mathcal{I}_0(nk_cb)}  \sin(n \zeta). \hspace{0.2in}
\end{eqnarray}

The magnetic field in the $\hat{z}=\hat{\zeta}$ direction in the absence of any currents between the poloidally-encircling coils is 
\begin{eqnarray}
B_z &=& \frac{ \partial \phi_{in}}{\partial z}= k_c \frac{ \partial \phi_{in}}{\partial \zeta}  \\
&=&\frac{ \mu_0G_c k_c}{2\pi} \Big(1 \nonumber\\
&&+2\sum_{n=1}^\infty (-1)^n \frac{I_0(nk_c r)}{\mathcal{I}_0(nk_cb)}  \cos(n \zeta) \Big).
\end{eqnarray}
After multiplying by the modified Bessel function factor $I_0(nk_c r)/\mathcal{I}_0(nk_c b)$ the convergence is rapid except near $\zeta=\pm\pi$ for $r$ significantly less than $b$, the smaller $r/b$ the more rapid the convergence.  

H-mode tokamaks are sensitive to the ripple near the plasma edge.  Stellarator power plants often have a relatively cold plasma near the edge, which would make the relevant $r/b$ smaller at the place where the acceptable ripple should be calculated.  The central ripple in the $n^{\mbox{th}}$ Fourier term is much smaller, $2/\mathcal{I}_0(nk_cb)$.

The asymptotic expansions of modified Bessel functions are adequate for most calculations:
\begin{eqnarray} 
I_0(x \rightarrow \infty) &=& \frac{e^x}{\sqrt{2\pi x}} \\  
K_0(x \rightarrow \infty) &=& \frac{e^{-x}}{\sqrt{2\pi x}} \\ 
\mathcal{I}_0(x \rightarrow \infty) &=& 2\frac{e^x}{\sqrt{2\pi x}}\\
B_z(r,\zeta) &=&\frac{ \mu_0G_c k_c}{2\pi} \Big(1 \nonumber\\
&&+\sqrt{\frac{b}{r}} \sum_{n=1}^\infty (-1)^n e^{-nk_c(b-r)}  \cos(n \zeta) \Big).  \label{Rippled field} \nonumber \\
\end{eqnarray}

When the only coils are $N_c$ poloidally encircling coils, then $k_c=N_c/R_0$ with $R_0$ the major radius.  To make the ripple less that 0.1\% at the plasma edge, $r=a$ the required number of coils would be 
\begin{equation}
N_c\approx 7\frac{R_0}{b-a}. \label{Coil number}
\end{equation}  

The W7-X stellarator has a major radius $R_0=5.5~$m, a plasma radius $a=0.53$~m, which gives the fifty modular coils that W7-X has if the coil radius were $b=1.3~$m.  The coils of W7-X are far from circular but have a half-height of 1.75~m.  Using that number for $b$ would give 32 coils.

The space between the poloidally encircling coils---the toroidal or modular field coils---can be increased by a factor $n_c$ by using localized coils to cancel the $n\leq n_c$ terms in Equation (\ref{Rippled field}) for $B_z$.  Equivalently, a single-valued current potential is added between the encircling coils with
 \begin{eqnarray}
 \kappa(\zeta) &=&  \frac{G_c}{\pi }\sum_{n=1}^{n_c}\frac{(-1)^n}{n} \sin n\zeta. \label{nulling kappa}
 \end{eqnarray}  

 When $n_c= 5$, W7-X with only two coils per period would have the equivalent ripple of the existing ten coils per period.  With $n_c=3$ and three coils per period, the ripple in W7-X would be only slightly increased with the equivalent ripple of 45 coils.

Where the reduced number of coils should be optimally placed within each period of a stellarator requires study.  In Todd Elder's \cite{Elder:2024} study of dipole patches to produce the non-axisymmetric fields of HSX, the patches are concentrated where ripple reduction is most important---the outboard side---at the corners of periods, Figure \ref{Patch-cov}.  Whether patches in the same area can efficiently reduce the toroidal ripple as well remains to be determined, which would mean the required toroidal field or modular coils would be located elsewhere.   

Despite the importance of opening the access to the plasma chamber, shockingly little has be studied.



\section{Discussion}
Machine maintenance and modification strategies for both tokamaks and stellarators strain credibility and leave much to be desired.  The difficulty is in large part coil design.  This paper points out a low-cost computational strategy for determining what is possible.  Can large spaces be created between the toroidal field coils of tokamaks to allow the removal of large components from the plasma chamber?  A factor of $n_c$ increase in the separation can be achieved by adding current potential patches, Equation (\ref{nulling kappa}), that null the $n\leq n_c$ terms in Equation (\ref{Rippled field}) for $B_z$.  A schematic example for $n_c=3$ is illustrated in Figure \ref{fig:removable-coils}.

  Neutron damage will require frequent replacements of internal components.  The economics of fusion requires this be done with minimal downtime.  The development of power plant concepts would be expedited if tradeoffs between alternative solutions could be studied by making major changes in an experiment rather than building a new machine.  
  
  For stellarators the coil issue is not just access but also ease of coil construction.  The complex shaping of toroidal field coils into modulars introduces numerous issues.  But, there are alternatives, which need to be explored.

The allowable separation between the plasma and the coils that produce the external magnetic field is a fundamental determinant in the minimum size and therefore cost of a fusion device.  A large coil-plasma separation together with easy access would allow  multiple plasma shapes, divertor designs, and blanket concepts to be studied over the lifetime of a single device.  Again, there are alternative coils that can be cheaply, quickly, and definitively studied computationally.  

Although not discussed in this paper, because they were extensively treated in Reference \cite{Comp-needs}, are coils for plasma control and for the mitigation of the effects of construction errors.  Such coils could greatly reduce the required time and cost of machine construction while increasing the utility of the machine for studying a broad range of issues.  

\section*{Acknowledgements}

This material is based upon work supported by the Grant 601958 within the Simons Foundation collaboration Hidden Symmetries and Fusion Energy, and by the U.S. Department of Energy, Office of Science under Award Nos. DE-FG02-95ER54333, DE-FG02-03ER54696, DE-SC0024548, and DE-AC02-09CH11466. 
 \vspace{0.01in}

\section*{Author Declarations}

The author has no conflicts to disclose. \vspace{0.01in}


\section*{Data availability statement}

Data sharing is not applicable to this article as no new data were created or analyzed in this study.


\appendix

\section{Gaussian smoothing \label{Gaussian smoothing} }

When a surface is uniformly gridded with the grid cell located at $\vec{x}_\ell$ having a magnetic dipole of strength $d_\ell$, then when the area of each cell is $a_\ell$, the current potential at the location of the cell is $\kappa_\ell=d_{\ell}/a_{\ell}$.   The smoothed current potential $\bar{\kappa}(\vec{x})$ on the surface is defined using a Gaussian function $G(\vec{x}-\vec{x}_\ell,\sigma)$.  The gradient of the smoothed current potential $\vec{\nabla}\bar{\kappa}$ can be easily calculated and the smoothed values of the dipole strengths in each cell detrmined.

\begin{eqnarray}
G(\vec{x}-\vec{x}_\ell,\sigma) &\equiv& \frac{e^{-\frac{(\vec{x}-\vec{x}_\ell )^2}{2\sigma^2}}}{2\pi\sigma^2} ; \\
\bar{\kappa}(\vec{x}) &\equiv& \sum_\ell \kappa_\ell G(\vec{x}-\vec{x}_\ell,\sigma)  a_\ell \nonumber\\
&=& \frac{1}{2\pi\sigma^2} \sum_\ell  d_\ell  e^{-\frac{(\vec{x}-\vec{x}_\ell )^2}{2\sigma^2}};\\
\vec{\nabla}\bar{\kappa} &=&- \frac{1}{2\pi\sigma^2} \sum_\ell  \frac{\vec{x}-\vec{x}_\ell }{\sigma^2} d_\ell  e^{-\frac{(\vec{x}-\vec{x}_\ell )^2}{2\sigma^2}}; \hspace{0.3in}\\
\bar{d}_\ell  & \equiv & \bar{\kappa}(\vec{x}_\ell ) a_\ell. \label{smooth-d}
\end{eqnarray}
As in the rest of the paper, the cells will be assumed to have equal areas with $a_\ell=a_c$.  When the smoothing distance $\sigma$ goes to zero $\oint G da =  \delta(\vec{x}-\vec{x_\ell })$ is a delta function at the location of the cell. 

As will be shown in Appendix \ref{Gaussian-sigma}, the smoothing distance can be chosen as $\sigma=\sqrt{a_c}$ and obtain a good fit to $\bar{\kappa}$ with only a five term sum in the Gaussian smoothing.   As will be shown in  Appendix \ref{Gaussian-avg}, the difference between the smoothed and the actual current potential scales as $(k\sigma)^2$, where $k$ is the wavenumber of variations in the current potential.  When $k\Delta>2$, with $\Delta$ the plasma-coil separation, the field will drop be more than a factor of five between the coils and the plasma.  Consequently, $a_c/\Delta^2 = (k\sigma)^2/k\Delta\approx 1/200$ for one-percent accuracy.

With the symmetries of a standard stellarator, only a half period has independent coils, which usually has a length of approximately $\pi a$, where $a$ is the plasma radius.  The area of the wall $A_w \approx 2\pi^2 (a+\Delta)a \approx 4\pi^2 \Delta^2$ when the coil-plasma separation and the plasma radius are approximately equal.  The number of cells is $A_w/a_c \approx 40 (\Delta^2/a_c) \approx 8000$ for one-percent accuracy.

To simplify the discussion in Appendices  \ref{Gaussian-sigma} and \ref{Gaussian-avg}, one-dimensional Gaussian smoothing will be considered.  When the two coordinates of the surfaces are approximately orthogonal, the required two-dimensional smoothing is the product of two one-dimensional smoothing operations.


The Gaussian smoothing width $\sigma$ must be sufficiently large compared to the cellular discretization to accurately represent the current potential even in a region in which it is constant and must be sufficiently small compared to the wavenumber $k$ of the current potential $k^2\equiv \kappa"/\kappa$, where $\kappa"$ is the second spatial derivative  of the current potential in either of its coordinate directions. \color{black} \vspace{0.1in}

\subsection{Properties of a Gaussian series \label{Gaussian-sigma}}

If the $d_\ell$ were all equal, how large would $\sigma$ need to be to obtain an accurate representation of a constant current potential?   In a one-dimensional problem, the question would be how large would $w\equiv\sigma/\sqrt{a_c}$ need to be for $\mathcal{G}(w)$ to be essentially unity, where 
\begin{eqnarray} 
\mathcal{G}(w) &\equiv& \frac{1}{w\sqrt{2\pi}} \sum_{n=-\infty}^{\infty} \exp\Big(- \frac{1}{2}\left(\frac{n}{w}\right)^2\Big).
\end{eqnarray}
Our numerical calculation is given in the table
\begin{eqnarray}
\mbox{$\mathcal{G}$ versus $w$ including all $n$ } \\
\left(\begin{array}{cc} w & \mathcal{G} \\0.25 & 1.59683976 \\0.5 & 1.01438377 \\1/\sqrt{2} & 1.00010345 \\1.0 & 1.00000001 \\2.0 & 1.00000000 \\3.0 &1.00000000 \\4.0 & 1.00000000 \\ 5.0 & 0.99999901\end{array}\right) \\ \nonumber \\
\mbox{$\mathcal{G}$ versus largest retained $|n|$ for $w=1$}\nonumber\\\
\left(\begin{array}{cc} |n| & \mathcal{G} \\1 & 0.8828837 \\2 & 0.99086566 \\3 & 0.9997294 \\4 & 0.9999970 \end{array}\right)
\end{eqnarray}

The calculation of $\mathcal{G}$ was checked by comparing to a result given by WolframAlpha,
\begin{eqnarray}
\mathcal{G}\left(\frac{1}{\sqrt{2}}\right) = 1.00010345\cdots
\end{eqnarray}
to eight significant digits. \vspace{0.1in}


\subsection{Gaussian average of a function \label{Gaussian-avg}}

Although $\sigma$ needs to be sufficiently large for the discrete sum to accurately represent even a constant current potential, the smaller $\sigma$ the better the variations in the current potential are represented.  For simplicity, a current potential that depends on only one coordinate is considered.

\begin{eqnarray}
\bar{\kappa}(x) &\equiv& \frac{1}{\sigma\sqrt{2\pi}} \int_{-\infty}^\infty \kappa(y) e^{-\frac{(x-y)^2}{2\sigma^2}}  dy. \hspace{0.1in} \mbox{    Let   } \hspace{0.2in} \\
y &\equiv& x-\sigma t \hspace{0.1in} \mbox{    then   } \hspace{0.2in} \\
\bar{\kappa}(x) &\equiv& \frac{ \int_{-\infty}^\infty \kappa(x-\sigma t) \exp(-t^2/2) dt}{\int_{-\infty}^\infty \exp(-t^2/2) dt}
\end{eqnarray}

The integral of powers of $t$ multiplied by a Gaussian vanish for odd powers of $t$ by symmetry.  For even powers, they can be calculated by differentiation using
\begin{eqnarray}
I(\alpha) &\equiv& \int_{-\infty}^\infty \exp(-\alpha t^2/2) dt \\
&=& \alpha^{-1/2}  \int_{-\infty}^\infty \exp(- t^2/2) dt \\
\frac{dI}{d\alpha} &=& - \frac{1}{2}\alpha^{-3/2} \int_{-\infty}^\infty \exp(- t^2/2) dt, \hspace{0.1in} \mbox{    so   } \hspace{0.2in} 
\end{eqnarray}
\begin{eqnarray}
&& \int_{-\infty}^\infty t^2 \exp(- t^2/2) dt= \int_{-\infty}^\infty\exp(- t^2/2) dt. \hspace{0.2in} \\
&& \int_{-\infty}^\infty t^4 \exp(- t^2/2) dt = 3 \int_{-\infty}^\infty\exp(- t^2/2) dt. \hspace{0.3in}\\
&& \bar{\kappa}(x) = \kappa(x) + \frac{\sigma^2}{2} \kappa"(x) +  \frac{\sigma^4}{2}\kappa^{\mbox{iv}}(x) + \cdots \hspace{0.2in}
\end{eqnarray}



\begin{thebibliography}{99}

\bibitem{Comp-needs} A. H. Boozer, \emph{Needed computations and computational capabilities for stellarators}, Phys. Plasmas \textbf{31}, 060601 (2024).  

\bibitem{Merkel} P. Merkel, \emph{Solution of stellarator boundary value problems with external currents}, Nucl. Fusion, \textbf{27}, 867 (1987). 

\bibitem{ITER first wall} A. H. Boozer, \emph{The interaction of the ITER first wall with magnetic perturbations}, Nucl. Fusion \textbf{61}, 046025 (2021).  

\bibitem{ITER hot Cell: 2017} J-P Friconneau, V. Beaudoin, A Dammann, C. Dremel, JP Martins, and C.S. Pitcher, \emph{ITER hot Cell---Remote handling system maintenance overview}, Fusion Engineering and Design \textbf{124}, 673 (2017). 

 \bibitem{Modulars:1972} H. Wobig and S. Rehker, \emph{A Stellarator Coil System without Helical Windings}, Proc. 7th
Symp. on Fusion Technology, Grenoble, France, 333-343 (October 24-27, 1972), \scriptsize $<$https://op.europa.eu/en/publication-detail/-/publication/761661a8-8c1d-4959-bc59-dcdc29acfa94$>$. \normalsize

\bibitem{Focus code: 2017} C. Zhu, S. R. Hudson, Y. Song, and Y. Wan, \emph{New method to design stellarator coils without the winding surface}, Nucl. Fusion \textbf{58}, 016008 (2017). 

\bibitem{Boozer:RMP} A. H. Boozer, \emph{Physics of magnetically confined plasmas}, Rev. Mod. Phys., \textbf{76}, 1071 (2004). 

 \bibitem{Landreman:2017} M. Landreman, \emph{An improved current potential method for fast computation of stellarator coil shapes},  Nucl. Fusion \textbf{57} 046003 (2017). 

 \bibitem{Zhu et at} C. Zhu, M. Zarnstorff, D. Gates, and A. Brooks, \emph{Designing stellarators using perpendicular permanent magnets}, Nucl. Fusion \textbf{60}, 076016 (2020). 
 
 \bibitem{Hammond:2024} K. C. Hammond and A. A. Kaptanoglu, \emph{Improved stellarator permanent magnet designs through combined discrete and continuous optimizations}, Computer Physics Communications \textbf{299},  109127 (2024). 
 
\bibitem{Thea Energy:2023} D. Gates, S. Aslam, B Berzin, A. Cote, D. W.  Dudt, D. Fort, A. Koen, T. Kruger, S. T. Kumar, M. F. Martin, C. P. S. Swanson, and E. Winkler, \emph{Thea Energy: Reinventing the stellarator}, Bulletin of the American Physical Society, 65th Annual Meeting of the APS Division of Plasma Physics, October 30-November 3 2023, Abstract: CM06.00010.

\bibitem{Kaptanoglu:2024a}  A. A. Kaptanoglu, G. P. Langlois, M. Landreman, \emph{Topology optimization for inverse magnetostatics as sparse regression: Application to electromagnetic coils for stellarators}, Comput. Methods Appl. Mech. Engrg. \textbf{418}, 116504 (2024). 



\bibitem{Landreman:eff-B} M. Landreman and A. H. Boozer, \emph{Efficient magnetic fields for supporting toroidal plasmas}, Phys. Plasmas \textbf{23}, 032506 (2016). 

\bibitem{Landreman:field-gradient} J. Kappel, M. Landreman, and D. Malhotra, \emph{The Magnetic Gradient Scale Length Explains Why Certain Plasmas Require Close External Magnetic Coils}, Plasma Phys. Control. Fusion, \textbf{66}, Issue 2 (2023). 


\bibitem{Ku-Boozer:2010} L.-P. Ku and A. H. Boozer, \emph{Stellarator coil design and plasma sensitivity}, Phys. Plasmas \textbf{17}, 122503 (2010). 

\bibitem{Boozer-Ku:control2011} A. H. Boozer and L. P. Ku, \emph{Control of stellarator properties illustrated by a Wendelstein7-X equilibrium}, Phys. Plasmas \textbf{18}, 052501 (2011). 

\bibitem{Boozer:2009} A. H. Boozer, \emph{Use of nonaxisymmetric shaping in magnetic fusion}, Phys. Plasmas \textbf{16}, 058102 (2009). 

\bibitem{Elder:2024} T. Elder and A. H. Boozer, \emph{Current potential patches}, $<$ https://arxiv.org/pdf/2402.17634 $>$.

\bibitem{HSX} B. Anderson, F. Simon, A. F. Almagari, D. T. Anderson, P. G. Matthews, J. N. Talmadge, and J. L. Shohet, \emph{The helically symmetric experiment, (HSX) goals, design and status}, Fusion Technol. \textbf{27} (3T), 273 (1995). 


\bibitem{Kaptanoglu:2024b} A. A. Kaptanoglu, R. Conlin, and M. Landreman, \emph{Greedy permanent magnet optimization}, Nucl. Fusion \textbf{63},  03601 (2023). 



\bibitem{Cerfon} D. Malhotra, A. J, Cerfon, M. O'Neil, and E. Toler, \emph{Efficient high-order singular quadrature schemes in magnetic fusion}, Plasma Phys. Control. Fusion \textbf{62} 024004 (2020). 




 


\end{thebibliography}
\end{document}